\documentclass[aps,twocolumn,showpacs]{revtex4}
\usepackage{graphicx}
\begin{document}
%\draft

\flushbottom
%\twocolumn[
%\hsize\textwidth\columnwidth\hsize\csname @twocolumnfalse\endcsname

\title{Power loss in open cavity diodes and a  modified Child Langmuir Law }
\vskip 0.3 in
\author{Debabrata Biswas}
\author{Raghwendra Kumar}
\affiliation{Theoretical Physics Division,
Bhabha Atomic Research Centre,
Mumbai 400 085, INDIA}
\email{dbiswas@barc.ernet.in, raghav@barc.ernet.in }

\author{R.~R.~Puri}
\affiliation{Human Resource Development Division, 
Bhabha Atomic Research Centre, 
Mumbai 400 085, INDIA}
\email{rrpuri@barc.ernet.in}
\date{\today}

\vskip 0.2 in
%\centerline{\bf Abstract}
\begin{abstract}
Diodes used in most high power devices are inherently open. 
It is  shown that under such circumstances, there is a 
loss of electromagnetic radiation leading to a lower 
critical current as compared to closed diodes. 
The power loss can be incorporated in the standard Child-Langmuir
framework by introducing an effective potential. The modified 
Child-Langmuir law can be used to predict the maximum power
loss for a given plate separation and potential difference
as well as the maximum transmitted current for this power loss.
The effectiveness of the theory is tested numerically.  

\vskip 0.25 in
\end{abstract}
\maketitle

\date{today}
%]
%\narrowtext

%date{today}
%\end{frontmatter}
\newcommand{\be}{\begin{equation}}
\newcommand{\ee}{\end{equation}}
\newcommand{\bea}{\begin{eqnarray}}
\newcommand{\eea}{\end{eqnarray}}
\newcommand{\Tbar}{{\overline{T}}}
\newcommand{\En}{{\cal E}}
\newcommand{\Lop}{{\cal L}}
\newcommand{\DB}[1]{\marginpar{\footnotesize DB: #1}}
\newcommand{\q}{\vec{q}}
\newcommand{\kt}{\tilde{k}}
\newcommand{\Lopn}{\tilde{\Lop}}

\newpage

\section{Introduction}
\label{sec:intro}

       The motion of charged particles accelerated
across a gap is of considerable interest in fields such as high power
diodes and vacuum microelectronics.
There are two possible mechanisms that lead to a saturation
in the transmitted current. One, where the voltage difference 
is increased till all the charges that are produced at the cathode,
are transmitted.
The other, where the voltage is held fixed and the charges produced at the 
cathode are made to increase (e.g. by increasing the temperature 
in a thermionic diode or by increasing the power of a laser
in case of a photocathode). Interestingly, a saturation is observed 
even in the second case and the current is said to be space
charge limited. A useful approximation for the amount of current flowing
in such cases is the Child-Langmuir expression for transmitted 
current between two infinite parallel plates.
For electrodes having a potential difference $V$ and separated by
a distance $D$, the Child-Langmuir current density is 
expressed as \cite{child,langmuir,BB}

\be
J_{\rm CL}(1) = {V^{3/2} \over D^2} {1 \over 9 \pi}
\left({2e\over m_0}\right)^{1/2}.
\label{eq:CL_1d}
\ee

\noindent
where $e$ is the magnitude of the charge and $m_0$ the rest mass of 
an electron. In one-dimensional situations (plate dimension much larger
than their separation), this is also the current density at
which some electrons, instead of moving towards the anode, return to
the cathode due to mutual repulsion (space charge effect). This
is referred to as critical current density. In the above
scenario, the initial velocity of the electrons is zero. Generalizations
for non-zero initial velocity can be found in \cite{non0,akimov,pop3}.

In  two dimensions, analytical and
numerical studies \cite{lugins,CL2D,lau} 
indicate that the critical current is somewhat
higher than the Child-Langmuir expression and is given by \cite{lugins,lau}

\be
{J_{\rm CL}(2) \over J_{\rm CL}(1)} = 1 + a_0 {D \over W}.
\label{eq:CLsquare}
\ee

\noindent
for infinite plates with a finite emission area.
Here $W$ is the width of the emission strip, $D$ is the anode-cathode
separation and $a_0 \simeq 0.31$. For a circular emission area 
of radius R, the Child-Langmuir law takes
the form \cite{lau}

\be
{J_{\rm CL}(2) \over J_{\rm CL}(1)} = 1 + {D \over 4R} \label{eq:CLcirc}.
\ee

\noindent
in the limit $R/D >> 1$.

	The picture presented above is considerably simplified
as it is based on electrostatic analysis.
Nonetheless, it suffices when the diode is closed. In reality, as
the current starts flowing, electromagnetic radiation is emitted.
In a closed cavity, the emitted radiation opposes the flow of 
electrons in the region near the cathode while in the region 
towards the anode, it enhances the acceleration due to the applied 
field \cite{fnote0}. Thus, electrons regain the emitted energy from
the electromagnetic field so that the net field energy attains
equilibrium.

Most often
however, diodes are ``open'' due to the presence of either
a mesh anode (such as in a vircator) or insulating dielectrics
in the pulse forming line  of the cavity diode or a dielectric
window as in a photocathode. 
This results in a loss of electromagnetic radiation.  The 
electrons are thus unable to reabsorb all the emitted radiation
and hence acquire a lower energy on reaching the anode as compared
to the case of a closed diode. This also leads to a drop in
transmitted current due to the enhanced repulsion between the
slowly moving electrons. Thus, diodes open to electromagnetic
radiation cannot be governed by any of the above Child-Langmuir
laws. 
Interestingly, there are several instances where deviations 
from the Child-Langmuir law {\em have} been 
observed \cite{chung,nassisi02,nassisi94}.
In cases where the current measured is considerably higher, modifications
to the Child-Langmuir law take into account the expansion
of the plasma formed at the cathode. This leads to a decrease 
in the anode-cathode 
separation \cite{chung,nassisi02} with time, $t$. 
The separation $D$ in eq.~(\ref{eq:CL_1d}) is thus
replaced by $D - vt$ where $v$ is the plasma 
expansion rate \cite{nassisi02}. Transmitted currents lower
than the Child-Langmuir law have also been observed.
On completion of this work, we were made aware of such
a case in experiments using photocathode. An explanation put forward for this 
observation takes into account the impedance of the plasma formed
at the cathode so that the voltage $V$ in  eq.~(\ref{eq:CL_1d})  
is replaced by $V - ZI$ where $I$ is the current ($J A$ where 
$A$ is the emitter area)
and $Z$ is the plasma impedance \cite{nassisi94,nassisi02}. 
However, comparisons with actual measurements of impedance
were not made and the problem may thus be considered open.
The form of this modified law (replacing 
$V$ by $V - ZI$) is however 
general enough to account for any loss mechanism and the one 
that we propose here due to 
leakage of electromagnetic radiation can also be expressed 
in such a form. While we are not aware of direct
measurements of power loss in a diode, it is our hope nonetheless that the
modification that we propose serves to explain some of the deviations
from the standard Child-Langmuir law observed in experimental situations.

We shall first demonstrate in section \ref{sec:simulation}
that there is indeed 
a leakage of electromagnetic radiation and a drop in transmitted current
in open diodes using a Particle In Cell (PIC) simulation. 
We shall thereafter show (section \ref{sec:modified_CL}) 
how the Child-Langmuir can be
modified to account for the loss of energy due to the leakage
of electromagnetic radiation. 

\section{PIC simulation of Open Diode}
\label{sec:simulation}

\begin{figure}[tbp]
\begin{center}
\includegraphics[width=8cm,angle=0]{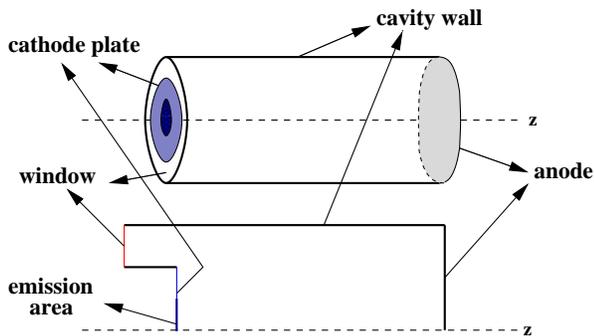}
\end{center}
\caption[ty] {(Color online) A schematic of the diode. 
The dark patch on the cathode plate is the emission area. 
The figure at the bottom 
shows the diode in z-r plane as modelled in XOOPIC.}
\label{fig:1}
\end{figure}

Consider a cavity diode as shown in fig.~1. It consists of a
cathode plate with emission from a part of the cathode plate (the dark 
region in fig.~1), an
anode plate and an outer cavity wall. The gap between the 
cathode plate and the cavity wall is the ``open window''
through which radiation leaks out. We shall consider here
a cathode-anode separation, $D = 8$ mm, the radius of the 
cavity wall $R_{W} = 12.5$ cm, the radius of the circular emission
strip $R = 3.5$ cm and the applied voltage on the cathode-plate 
$V = -250$ kV (the other parts of the diode are grounded). The 
radius of cathode plate, $R_{CP}$ varies from 4 cm to 12.5 cm, the last
corresponding to a closed diode. For the closed geometry, eq.~(\ref{eq:CL_1d})
predicts $J_{\rm CL}(1) \simeq 4.57$ MA/${\rm m}^2$ while 
eq.~(\ref{eq:CLcirc})
gives $J_{\rm CL}(2) \simeq 4.83$ MA/${\rm m}^2$.
For a beam radius of $3.5$ cm, the corresponding currents are
$17.58$ KA and $18.58$ KA respectively.

\begin{figure}[tbp]
\begin{center}
\includegraphics[width=6cm,angle=270]{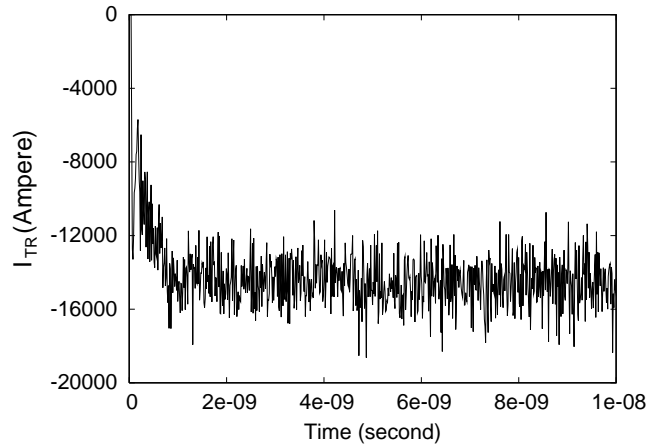}
\end{center}
\caption[ty] {The transmitted current shows a steady state behaviour after
initial transients. Here $R_{CP} = 0.12$ m. }
\label{fig:steady_state}
\end{figure}

Our numerical studies have been carried out using the 
fully electromagnetic particle-in-cell 
(PIC) code XOOPIC version 2.51 b(c) \cite{xoopic} 
and verified using a suitably 
modified form of the electromagnetic PIC code SPIFFE 2.3 \cite{spiffe}. 
The modification in SPIFFE takes 
into account open boundaries by applying first order Mur type 
boundary conditions \cite{Mur}.

The results we present here have been obtained using XOOPIC.
In the input file, the cavity wall and anode plate are 
considered as ``conductors'' while the cathode plate at 
the ``beam emitter (VarWeightBeamEmitter or FieldEmitter)'' 
is an ``equipotential'' where the applied voltage is fixed
in time. The simulation begins with an evaluation of the static
fields by solving the Laplace equation. In order to study
more realistic voltage time profiles (figures 10 and 11), we
have suitably modified XOOPIC such that the applied fields
are evaluated at every time step. Unless otherwise stated,
the applied fields will be considered static in this paper.  

At the open window, an ``exitport'' is used to evaluate the
power emitted through it. The number of mesh points in
azimuthal and radial directions are typically 100 and 250 
while the time step $\Delta t = 2\times 10^{-13}$s.
The number of particles emitted per time step varies between 
200 and 700 depending on the current. These calculations
have been performed in the absence of any externally applied
magnetic field. 

For each opening, we increase the injected current till the
transmitted current attains (quasi) saturation \cite{fnote1,pop1}. 
We then record the transmitted current and the power emitted 
through the open window. Note that for each simulation, we
ensure the onset of steady state behaviour in the transmitted
current (see fig.~\ref{fig:steady_state})  to eliminate the effect 
of transients. 

\begin{figure}[tbp]
\begin{center}
\includegraphics[width=6cm,angle=270]{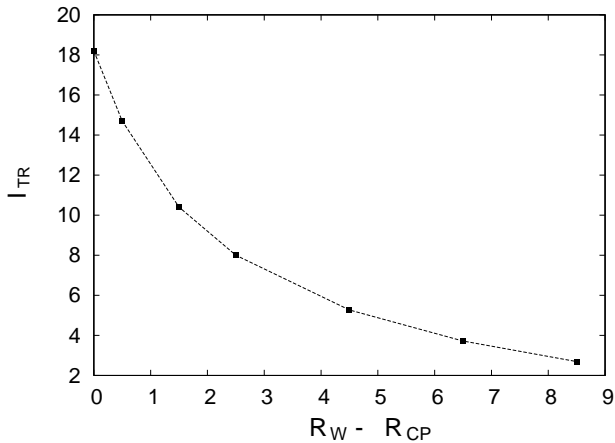}
\end{center}
\caption[ty] {The transmitted current $I_{TR}$ (KA) is plotted
against the opening $R_W - R_{CP}$ (cm). The transmitted current
decreases monotonically as the opening is increased. }
\label{fig:3}
\end{figure}

\begin{figure}[tbp]
\begin{center}
\includegraphics[width=6cm,angle=270]{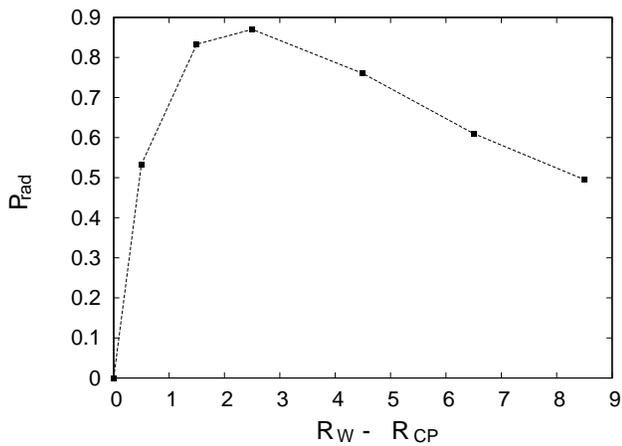}
\end{center}
\caption[ty] {The emitted power $P_{rad}$ (GW) as a function of
the opening  $R_W - R_{CP}$ (cm). The power peaks at $R_W - R_{CP} = 2.5$ cm.}
\label{fig:4}
\end{figure}

Fig. \ref{fig:3} shows a plot of the transmitted current density 
as a function of
the opening $R_W - R_{CP}$. There is a monotonic decrease as the
window opening increases. In figure \ref{fig:4}, we plot the power emitted
through the window as a function of the opening $R_W - R_{CP}$.
The power peaks at $R_W - R_{CP} = 2.5$ cm or $R_{CP} = 10$ cm.

To understand the relation between the area of the open window
and the emitted power, it should first be noted that 
the emitted power is zero for a closed diode ($R_W - R_{CP}= 0$)
even though the electromagnetic field energy inside the diode
is high. As the opening increases, some of this field energy
leaks out. Thus the emitted power increases, the current drops
and the field energy inside reduces as the opening is increased. 
However, at very large openings, most of the emitted radiation
leaks out. Consequently, the transmitted current and hence the
field energy inside the diode must be
small. Thus, even though the window area is large, the power emitted
must be small. It follows that at a critical value 
of the opening, the field energy 
inside, transmitted current and window area are such that
the power emitted is at a maximum as observed in fig.~\ref{fig:4}.

We thus find that the transmitted current decreases with an
increase in window area while the emitted power peaks
as the window area is increased from zero and then
decreases for larger openings \cite{load}. An exact theory predicting 
such a behaviour is rather formidable and shall not be
attempted here. However, in the following, we shall argue 
how the power loss can be incorporated within the Child-Langmuir
framework.

\section{Modified Child-Langmuir Law for open diodes}
\label{sec:modified_CL}

In a closed geometry, the average power transferred to the
electrons is $I_{TR} V$ where $I_{TR}$ is the average transmitted
current. The leakage of electromagnetic radiation through the
open window effectively results in a power loss. If $P_{rad}$
is the radiative power loss through the window, the effective
average power transferred to the electrons is $I_{TR} V - P_{rad}$.
Alternately, one may imagine that the diode is closed but
the effective applied voltage is 

\be
V_{eff} = V - P_{rad}/I_{TR} \label{eq:effV}.
\ee

\begin{figure}[tbp]
\begin{center}
\includegraphics[width=6cm,angle=270]{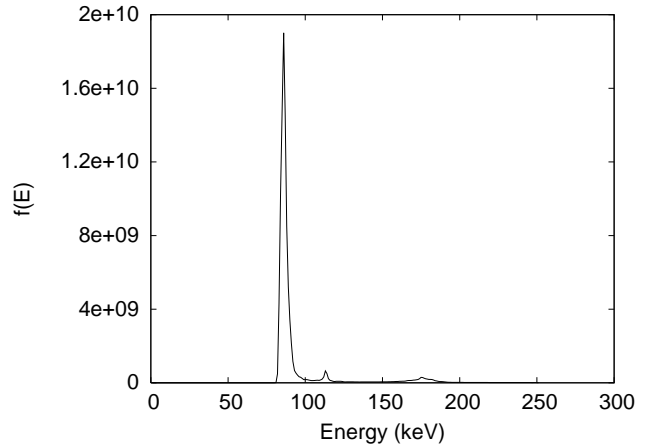}
\end{center}
\caption[ty] {The energy distribution, $f(E)$, of electrons reaching
the anode for $R_{CP} = 0.06$ m. Parameters such as the $V$, $D$ and $R$ are
as in previous plots. 
}
\label{fig:energy_stat}
\end{figure}

\begin{figure}[tb]
\vskip -0.15in
\begin{center}
\includegraphics[width=6cm,angle=270]{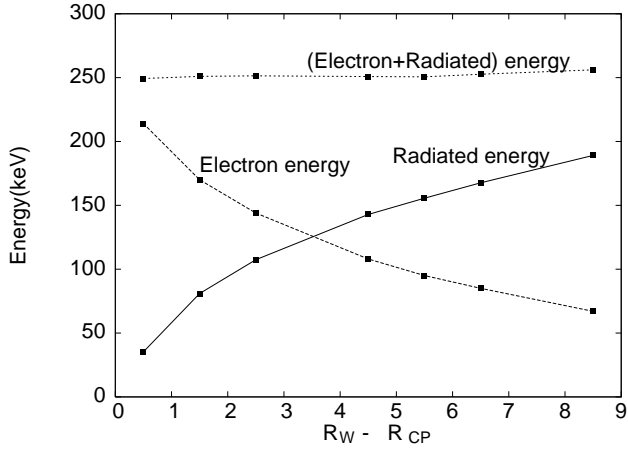}
\end{center}
\caption[ty] {The radiated energy ($e P_{rad}/I_{TR}$) and 
the electron energy are plotted 
separately for different window openings ($R_{W} - R_{CP}$ (cm)). The sum of
these energies is also plotted and lies around the 250 keV line. This 
is the energy that a electron would gain in a closed diode where the
radiated energy is zero. 
}
\label{fig:energy_sum}
\end{figure}

In the scenario presented here, the average energy of the 
electrons reaching the anode plate is $eV_{eff}$.
In other words, electrons give up some of their
energy in the form of radiation. In case of open diodes, this energy 
leaks out through the window, while, in closed diodes, this is reabsorbed 
by the electrons. As a check, we have computed the 
electron beam energy statistics 
at the anode. As the window opening increases, the energy of 
electrons reaching the
diode reduces. A typical plot of the energy distribution of electrons reaching
the anode for a particular window opening is 
shown in fig. \ref{fig:energy_stat}.
Clearly, the peak is sharp. The sum of the energy corresponding to this
peak and the radiated energy ($e P_{rad}/I_{TR}$), is found to 
be equal to the 
energy attained by electrons in a closed diode with the same 
applied potential $V$.
Fig.~\ref{fig:energy_sum} shows this for different window openings when the
applied voltage is $250$ kV.

We can now apply the usual Child-Langmuir analysis to the diode 
and obtain an expression for the current $I_{TR}$ in terms of the effective
voltage

\be
I_{TR} = {A~V_{eff}^{3/2} \over D^2} {1 \over 9 \pi}
\left({2e\over m_0}\right)^{1/2} 
\ee

\noindent
where $A = \pi R^2$ is the emission area. One can even improve upon
this one-dimensional calculation by using a geometric factor to account 
for two-dimensional effects
together with the
effective voltage to account for the finite emission area.
We thus have an expression for the transmitted current

\be
I_{TR} = \alpha_{g} \left({2e\over m_0}\right)^{1/2} {1 \over 9 \pi} 
{A~(V - P_{rad}/I_{TR})^{3/2} \over D^2}
\label{eq:modCL}
\ee

\noindent
where $\alpha_g$ is a geometric factor defined 
by $J_{CL}(2) = \alpha_g J_{CL}(1)$.
In practice, $\alpha_g$ should be determined numerically by measuring
$J_{CL}(2)$ for a closed diode and using eq.~(\ref{eq:CL_1d}) 
for $J_{CL}(1)$.    
Equation~\ref{eq:modCL} is our proposal for the transmitted 
current in open diodes
and forms the central result of this paper.

\begin{figure}[tbp]
\begin{center}
\includegraphics[width=8cm,angle=0]{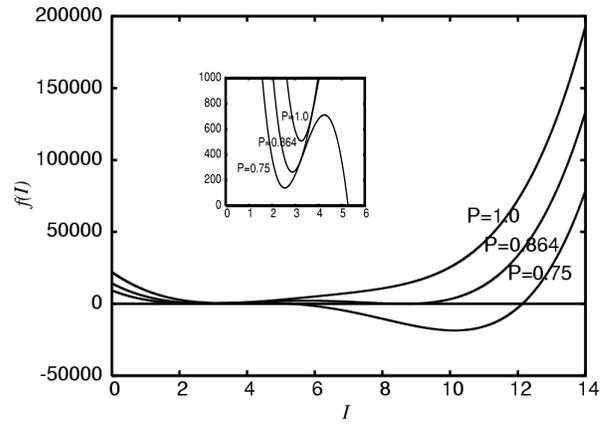}
\end{center} 
\caption[ty] {The function $f(I)$ in eq.~(\ref{eq:CLmod2}) plotted
as a function of the transmitted current $I$ for three different
values of the emitted power $P$. For $P = 0.75$, there are
2 positive roots while at $P = 0.864$ the two positive roots
merge. For $P > 0.864$ there are no positive roots. The inset shows
a blow up of the portion where the three curves appear to meet. None
of the curves however crosses the x-axis.}
\label{fig:7}
\end{figure}

Note that eq.~(\ref{eq:modCL}) does not allow us to read 
off the transmitted
current given the potential difference, plate separation and 
area of the open window. Rather, it depends on the power of the
radiation emitted through the open window, a quantity that is
unknown {\it a priori}. However, eq.~(\ref{eq:modCL}) does contain
relevant information in the form of the maximum power emitted 
through the window and the transmitted current at this power 
(or opening) as we show below. 

For the geometry under consideration, eq.~(\ref{eq:modCL}) takes
the form

\be
I_{TR}^5 =  1.9799 \times 10^4~\alpha_g^2~(VI_{TR} - P_{rad})^3
\label{eq:CLmod1}
\ee

\noindent
where $P_{rad}$ is measured in gigawatts, V in megavolts and $I_{TR}$ in
kiloamperes. The number of allowed solutions of eq.~(\ref{eq:CLmod1}) 
depends on the power emitted. In fig.~\ref{fig:7}, we plot 

\be
f(I) = I^5 - 1.9799 \times 10^4~\alpha_g^2~(VI - P)^3 
\label{eq:CLmod2} 
\ee

\noindent
with $\alpha_g = 1.057$ and $V = 0.25$ MV for three distinct 
values of the power $P$. For $P$ less than the critical power 
$P_{c} = 0.864$ GW,
there are two positive roots. These roots merge at $P = P_{c}$
while for $P > P_{c}$, there is no positive root. Thus, there
is a limit on the maximum power that can be emitted through the
window and the corresponding transmitted current can be obtained.

\begin{figure}[tbp]
\begin{center}
\includegraphics[width=5cm,angle=270]{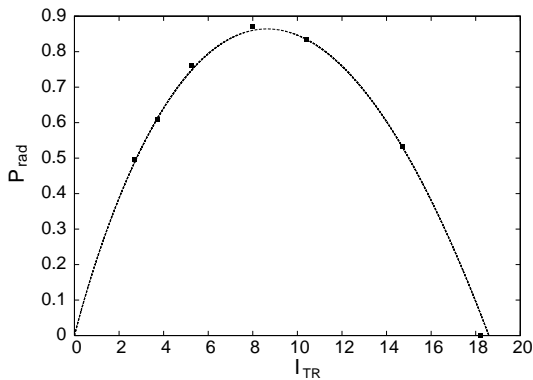}
\end{center} 

\caption[ty] {A comparison of the power (GW) leaking through the window
as a function of the transmitted current (KA) as calculated using XOOPIC 
(solid squares) with the modified Child-Langmuir law of eq.~(\ref{eq:modCL}).
for diode parameters as in fig.~3 and fig.~4. Here $D/R = 0.2285$.
}
\label{fig:8}
\end{figure}

%\vskip .25 in
%\newpage

\begin{figure}[tbp]

\begin{center}
\includegraphics[width=6cm,angle=270]{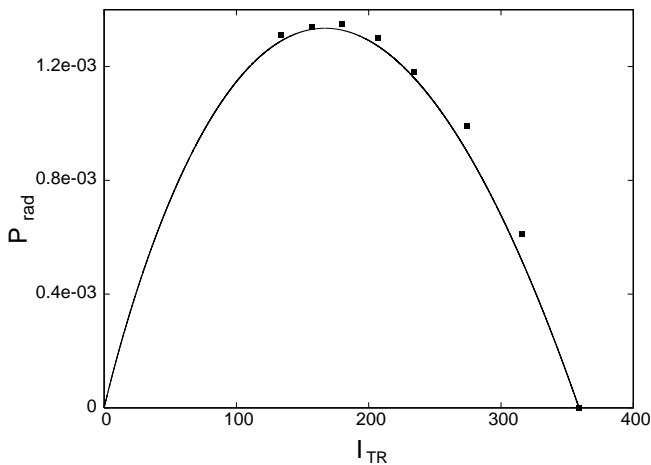}
\end{center} 
\caption[ty] {As in fig.~8 for a diode with cavity radius $0.1$ m, 
plate separation $D=0.005$ m, radius
of emitting area $R=0.02$ m ($D/R = 0.25$) and difference in plate potential 
$V = 20000$ volts.
The solid squares are simulation results using XOOPIC
while the dashed line is the prediction of eq.~(\ref{eq:modCL}) with
$\alpha_g$ determined numerically. 
}
\label{fig:9}
\end{figure}

In figure \ref{fig:8}, we plot the power emitted as a function of 
the transmitted
current as obtained numerically using XOOPIC. These are denoted 
by solid squares. The dashed curve is a plot of the same using 
the modified Child-Langmuir law of eq.~(\ref{eq:modCL}) with
$\alpha_g = 1 + D/4R$.  The 
agreement between the two is good. Note that
the emitted power is zero either when the diode is closed and the
current is maximum or when the current is zero.

We have used different $D/R$ ratios
and also different
voltages to study the Child-Langmuir law modified due to leakage
of electromagnetic radiation.
The agreement with eq.~(\ref{eq:modCL}) is good in general. 
An example where both the voltage and geometry are different from
the previous case is presented in fig.~9. 
Here $D = 0.005$ m, $R = 0.02$ m, $V = 20000$ V and 
the radius of the cavity wall is $0.1$ m. Note that $D/R = 0.25$. 
The prediction of eq.~(\ref{eq:modCL}) is again reasonable with 
$\alpha_g$ determined numerically using the closed diode. For 
comparison, the numerically determined $\alpha_g$ is found to be greater than
the prediction of eq.~(\ref{eq:CLcirc}) by a factor 1.02.

\begin{figure}[tbp]
\begin{center}
\includegraphics[width=6cm,angle=270]{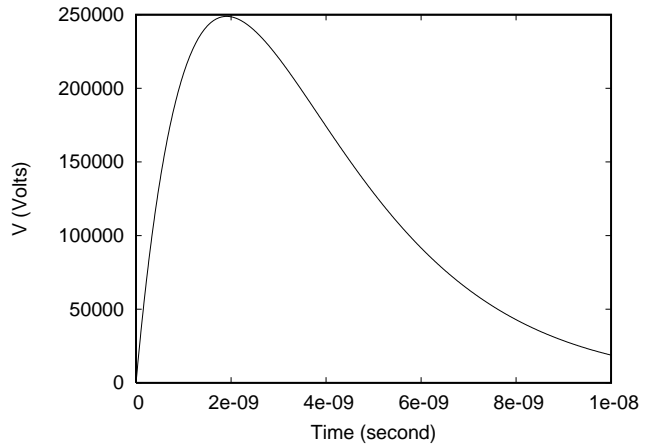}
\end{center} 
\caption[ty] {The time profile of cathode potential, $V$, used to study the frequency content
of the emitted radiation in fig.~\ref{fig:FFT}. 
}
\label{fig:potential}
\end{figure}

%\vskip .25 in
%\newpage

\begin{figure}[tbp]
\begin{center}
\includegraphics[width=7cm,angle=270]{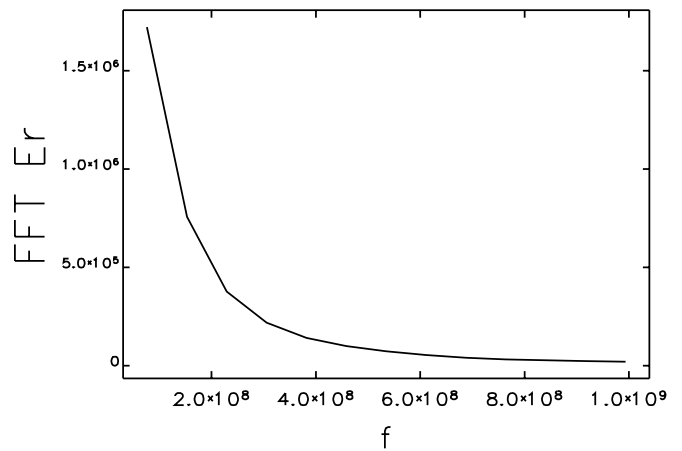}
\end{center} 
\caption[ty] {Fourier transform (FFT) of the radial component of electric field, $E_r$,
sampled near the exit window for the cathode potential shown in fig.~\ref{fig:potential}.
}
\label{fig:FFT}
\end{figure}

For completeness, a few remarks on the nature of the radiation
leaking out of an open diode are in order. If the potential at the
cathode is held constant, the emitted radiation is predominantly
in the low frequency end of the spectrum. In more realistic situations,
the time profile of the cathode potential is characterized by a rise and 
fall time. In such cases, the emitted radiation 
essentially contains frequencies 
that contribute to the time profile of the applied cathode potential.    
As an example, for the potential profile shown in fig.~{\ref{fig:potential},
the frequencies contained in the emitted radiation are predominantly 
in the low and intermediate frequency range as shown in fig.~\ref{fig:FFT}.

Finally, we have also used a dielectric window (with low $\epsilon$
such as in materials like Perspex) followed by an exit port. This results
in a marginal drop in power emitted and the modified
Child-Langmuir agrees with our simulations.

\section{Conclusions}

We have thus shown that an opening in a diode (such as due to 
the presence of dielectrics) acts as an exit port
for the emitted radiation and results in a drop in the
transmitted current. The power loss can be accomodated within
the Child-Langmuir framework by introducing an effective voltage
and the resulting modified law is in excellent agreement with  
PIC simulations using the code XOOPIC. It is our hope that in
experimental situations where leakage of radiation can occur, 
an analysis in this new light helps to explain deviations from the
standard Child-Langmuir law.

\newcommand{\PR}[1]{{Phys.\ Rep.}\/ {\bf #1}}
\newcommand{\PRL}[1]{{Phys.\ Rev.\ Lett.}\/ {\bf #1}}
\newcommand{\PP}[1]{{Phys.\ Plasmas\ }\/ {\bf #1}}

%\begin{references}


\begin{thebibliography}{99}
\bibitem{fnote1} For open diodes studied here, the transmitted current
attains quasi-saturation with increase in injected current.
The minor deviations observed may be ascribed to convergence errors. 
For closed cavity diode, it has been demonstrated that the transmitted 
current saturates only asymptotically with injected current  \cite{pop1}.
\end{thebibliography}
\end{document}